\documentclass[lettersize,journal]{IEEEtran}
\usepackage{amsmath,amsfonts}
\usepackage{algorithmic}
\usepackage{algorithm}
\usepackage{array}
\usepackage[caption=false,font=normalsize,labelfont=sf,textfont=sf]{subfig}
\usepackage{textcomp}
\usepackage{stfloats}
\usepackage{url}
\usepackage{verbatim}
\usepackage{graphicx}
\usepackage{cite}
\usepackage{multirow}
\usepackage{tabularx}
\usepackage{xcolor}
\usepackage[normalem]{ulem}
\usepackage{mathtools, cuted}
\usepackage{lipsum, color}
\usepackage[font=small,skip=0pt]{caption}
\begin{document}

\title{Photonic Reconfigurable Accelerators for Efficient Inference of CNNs with Mixed-Sized Tensors}

\author{Sairam Sri Vatsavai,~\IEEEmembership{Student Member,~IEEE,} Ishan G Thakkar,~\IEEEmembership{Member,~IEEE}

\thanks{Manuscript received April 07, 2022; revised June 11, 2022; accepted July 05, 2022. This article was presented in the International Conference on Compilers, Architectures, and Synthesis for Embedded Systems (CASES) 2022 and appears as part of the ESWEEK-TCAD special issue.}
}


\maketitle

\begin{abstract}
Photonic Microring Resonator (MRR) based hardware accelerators have been shown to provide disruptive speedup and energy-efficiency improvements for processing deep Convolutional Neural Networks (CNNs). However, previous MRR-based CNN accelerators fail to provide efficient adaptability for CNNs with mixed-sized tensors. One example of such CNNs is depthwise separable CNNs. Performing inferences of CNNs with mixed-sized tensors on such inflexible accelerators often leads to low hardware utilization, which diminishes the achievable performance and energy efficiency from the accelerators. In this paper, we present a novel way of introducing reconfigurability in the MRR-based CNN accelerators, to enable dynamic maximization of the size compatibility between the accelerator hardware components and the CNN tensors that are processed using the hardware components. We classify the state-of-the-art MRR-based CNN accelerators from prior works into two categories, based on the layout and relative placements of the utilized hardware components in the accelerators. We then use our method to introduce reconfigurability in accelerators from these two classes, to consequently improve their parallelism, flexibility of efficiently mapping tensors of different sizes, speed and overall energy efficiency. We evaluate our reconfigurable accelerators against three prior works for the area proportionate outlook (equal hardware area for all accelerators). Our evaluation for the inference of four modern CNNs indicates that our designed reconfigurable CNN accelerators provide improvements of up to 1.8$\times$ in Frames-Per-Second (FPS) and up to 1.5$\times$ in FPS/W, compared to an MRR-based accelerator from prior work. 
\end{abstract}


\section{Introduction}
Convolutional Neural Networks (CNNs) have shown record-breaking performance for implementing various real-world artificial intelligence tasks such as image recognition, language translation, and autonomous driving \cite{1,2,3,4}. The ever-increasing complexity of CNNs has pushed for highly customized CNN hardware accelerators \cite{8}. Among them, silicon-photonic CNN accelerators have shown great promise to provide unparalleled throughput, ultra-low latency, and high energy efficiency \cite{9,10,11,17,13,14}. Typically, a silicon-photonic CNN accelerator consists of multiple Tensor Product Cores (TPCs) that perform multiple tensor products in parallel. Several TPC-based photonic CNN accelerators have been proposed in prior works based on various silicon-photonic devices, such as Mach Zehnder Interferometer (MZI) (e.g., \cite{16},\cite{Ramey2020SiliconPF},\cite{mzi2018}) and Microring Resonator (MRR) (e.g., \cite{13,14,15},\cite{tait2020photonic}). 

Among these TPC-based photonic CNN accelerators from prior work, the MRR-enabled TPC-based accelerators (e.g., \cite{tait2017,tait2020photonic,13,14,15}) have shown disruptive performance and energy efficiencies for processing CNN tensor products, due to the MRRs' compact footprint, low dynamic power consumption, and compatibility with cascaded Dense-Wavelength-Division-Multiplexing (DWDM). The MRR-enabled TPCs of these accelerators transform CNN tensor products into Vector Dot Products (VDPs) by decomposing the input tensors into vectors (1D tensors). The VDP operations are performed on the individual VDP Elements (VDPEs), which are the  main MRR-enabled hardware components in a TPC. Multiple VDPEs in a TPC can perform multiple VDP operations in parallel. The results of these VDP operations can be summed together (when needed) using a partial-sum (\textit{psum}) reduction network, which can be employed outside of the TPCs as part of the post-processing components of the CNN accelerator. The functioning of the TPCs and their constituent VDPEs in the ultra-high-speed photonic domain results in disruptive throughput for processing tensors. 

However, the existing MRR-enabled TPC-based accelerators are not efficient in processing modern CNNs with mixed-sized tensors, such as Xception \cite{xception} and MobileNetV1 \cite{mobilenet}. This is because these modern CNNs utilize depthwise separable convolutions in addition to the standard convolutions. Depthwise separable convolutions in a CNN employ reduced sized tensors compared to the standard convolutions, to reduce the overall computational load of processing the CNN. For instance, MobileNetV1 shows 8-9$\times$ reduction in its computational load with only 1\% accuracy drop \cite{mobilenet}. But the existing MRR-enabled TPCs and their constituent VDPEs have fixed sizes. Therefore, mapping the processing of the modern CNNs with mixed-sized tensors \cite{31} on such fixed-sized TPCs often leads to a low hardware utilization in the TPCs. This in turn diminishes the achievable performance and energy efficiency from such fixed-sized TPCs-based accelerators.
This is because the low hardware utilization incurs non-amortizable area and static power overheads while also idling away the opportunity for increasing the processing throughput.

To address this shortcoming, in this paper, we present a novel way of introducing reconfigurability in the MRR-enabled TPCs-based CNN accelerators, to enable efficient support for both depthwise separable convolutions and standard convolutions. To enable this reconfigurability, we invent reconfigurable VDPEs that employ MRR-based comb switches to allow re-aggregation of the CNN vectors (decomposed 1D tensors) to consequently enable dynamic resizing of the produced VDP results. Our evaluations show that our invented reconfigurable VDPEs can (1) substantially improve the hardware utilization, (2) enhance the flexibility of processing CNN tensors of various sizes, (3) improve the opportunities for parallel tensor processing, and (4) significantly enhance the energy efficiency, for CNN inference acceleration.

Our key contributions in this paper are summarized below:

\begin{itemize}
\item We review several state-of-the-art MRR-enabled TPC-based CNN accelerators from prior work and then classify the circuit-level TPC organizations used in these accelerators into two categories, namely AMM and MAM (Section \ref{sec3});
\item We perform the scalability analysis of TPCs of AMM and MAM categories to capture the inter-dependence of the maximum achievable TPC size, bit precision, and operating bit rate (Section III-B);
\item We map the processing of four state-of-the-art CNNs with mixed-sized tensors on the TPCs of AMM and MAM categories to evaluate the hardware utilization of the TPCs, to consequently establish the need for reconfigurability in the AMM and MAM-styled TPCs (Section \ref{need_for_reconfigurability});
\item We invent a novel reconfigurable structure for VDPEs, and utilize these VDPEs to modify the AMM and MAM TPCs, to render these TPCs with the capabilities of dynamic re-aggregation of vectors for adaptive resizing of the processed VDPs (Section \ref{reconfig});
\item We evaluate the performance of our designed reconfigurable MAM and AMM TPC organizations for the inference of four modern CNNs in terms of Frames-Per-Second (FPS), FPS/W, and compare it with three different AMM- and MAM-styled CNN accelerators from prior work, with area proportionate outlook for which we set the area of all our evaluated accelerators to be equal (Section \ref{Evaluation}).
\end{itemize}

\section{Preliminaries}\label{sec2}

\subsection{CNNs with Mixed-Sized Tensors}
\begin{figure}[h!]
  \centering
  \includegraphics[scale=0.42]{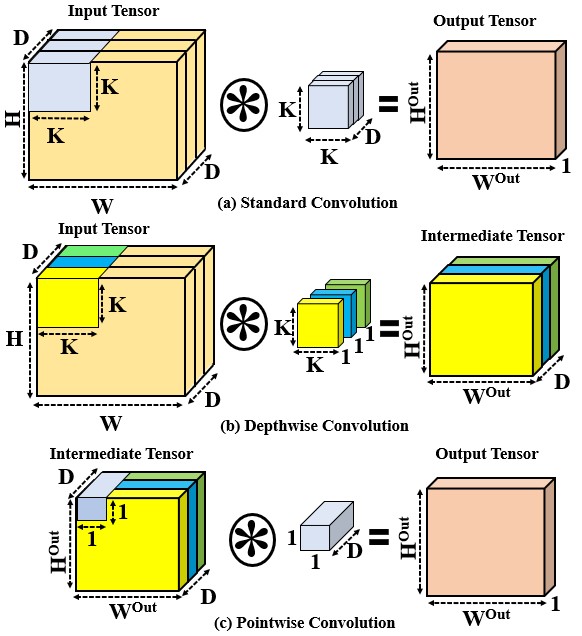}
  \caption{Illustration of various types of convolutions.}   
  \label{figdepthwiseconvo}
\end{figure}

It has been established that the efficiency of executing CNNs can be improved by drastically reducing the computation, communication, and memory requirements of the CNNs. To achieve this, there has been a growing trend of employing compressed, mixed-sized tensors in CNNs. For example, to compress the size of the utilized tensors, the Xception CNN model \cite{xception} introduced depthwise separable convolutions. Typically, a Depthwise Separable Convolution (DSC) breaks up a Standard Convolution (SC) into a Depthwise Convolution (DC) and a subsequent Pointwise Convolution (PC). Fig. \ref{figdepthwiseconvo} demonstrates how a Standard Convolution (SC), Depthwise Convolution (DC), and Pointwise Convolution (PC) work. An SC performs the channel-wise and spatial-wise tensor product computation in a single step by applying a single 3D kernel (convolutional filter) tensor to all the channels of the input tensor. In contrast, a DSC splits the tensor product computation into two steps. In the first step, the constituent DC applies a dedicated 2D kernel tensor per channel of the input tensor, and the channel-wise outputs are stacked to produce a single intermediate tensor. Then, in the second step, the subsequent PC is used to create a linear combination of all the channels of the intermediate tensor by applying a 1D kernel tensor for each spatial point of the intermediate tensor, to consequently produce the final output tensor. This can be better understood by referring to Fig. \ref{figdepthwiseconvo}, as explained next.

\subsubsection{Standard Convolutions (SCs)}
From Fig. \ref{figdepthwiseconvo}(a), in SC, the input tensor is \textit{H$\times$W$\times$D} in size, where \textit{H} is the height, \textit{W} is the width and \textit{D} is the number of channels (i.e., the depth). 
The kernel tensor has dimensions \textit{K$\times$K$\times$D}, which is convolved over the input tensor to generate a single output tensor with dimensions \textit{$H^{Out}\times W^{Out}\times 1$}. If there are F such kernel tensors convolved over the input tensor, then the output tensor will have the dimensions \textit{$H^{Out}\times W^{Out}\times F$} (Fig. \ref{figdepthwiseconvo}(a) shows F=1). If we define the tensor as a multidimensional array of points, then each point in the output tensor is obtained by performing a tensor product (sum of point-wise products) of the \textit{K$\times$K$\times$D} kernel tensor and the corresponding \textit{K$\times$K$\times$D} part of the input tensor (part of the input tensor highlighted in gray in Fig. \ref{figdepthwiseconvo}(a)). In this case, the total number of weight points across all of the F kernel tensors can be given by:

\begin{equation}
    W_{SC} = K\times K \times D \times F
\end{equation}

This $W_{SC}$ defines part of the memory and communication costs of generating the output tensor using SCs. Similarly, the total number of point-wise multiplication operations $O_{SC}$ (a point-wise multiplication between an input tensor point and a weight tensor point produces one point-wise product), required to generate all points in the output tensor, can be given by:

\begin{equation}
    O_{SC} = H^{Out} \times W^{Out} \times K\times K \times D \times F
\end{equation}

This $O_{SC}$ defines the computational cost of generating the output tensor using SCs.

\subsubsection{Depthwise Separable Convolutions (DSCs)}
As mentioned earlier, each DSC is typically split into a DC and a subsequent PC.  
From Fig. \ref{figdepthwiseconvo}(b), in DC, each channel of the input tensor has a corresponding 2D kernel tensor with dimensions \textit{K$\times$K$\times$1}. Since there are D channels in the input tensor, there are D such kernel tensors. Convolving these channel-wise kernel tensors across the spatial dimensions of their respective input channels generate a total of D channels of the intermediate tensor with each channel being of dimensions {$H^{Out}\times W^{Out}\times 1$}. Subsequently, in PC (Fig. \ref{figdepthwiseconvo}(c)), the intermediate tensor is convolved with a pointwise kernel tensor of dimensions \textit{1$\times$1$\times$D} to generate one output tensor of dimensions \textit{$H^{Out}\times W^{Out}\times 1$}. If there are F such point-wise kernel tensors, then the output tensor will have the dimensions \textit{$H^{Out}\times W^{Out}\times F$} (Fig. \ref{figdepthwiseconvo}(c) shows F=1). In this case, the total number of weight points across the DC and PC steps, considering F point-wise kernel tensors in the PC step, can be given by:

\begin{equation}
    W_{DSC} = K \times K \times D + D \times F
\end{equation}

Similarly, generating the final output tensor would require the total number of point-wise multiplication operations in the DCs, PCs, and DSCs to be $O_{DC}$, $O_{PC}$, and $O_{DSC}$, respectively, which can be given by:

\begin{equation}
     O_{DC} = H^{Out} \times W^{Out} \times K \times K \times D
\end{equation}
\begin{equation}
    O_{PC} =  H^{Out} \times W^{Out} \times D \times F
\end{equation}
\begin{equation}
     O_{DSC} = O_{DC}+O_{PC}
\end{equation}

Thus, for generating the final output tensor, the use of DSCs (DCs+PCs) results in a reduction in the required number of weight points and point-wise multiplication operations. This reduction can be given by the reduction factors $R_W$ and $R_O$:

\begin{equation}
    R_W = \frac{W_{DSC}}{W_{SC}} =  \frac{1}{F} + \frac{1}{K^2}
\end{equation}
\begin{equation}
    R_O = \frac{O_{DSC}}{O_{SC}} =  \frac{1}{F} + \frac{1}{K^2}
\end{equation}


It can be inferred that the use DSCs can reduce the memory+communication and computing costs of generating the output tensor by $R_W$ and $R_O$. Because of this advantage, several state-of-the-art CNN models, such as EfficientNet \cite{efficientnet}, ShuffleNetV2 \cite{shufflenet} and MobileNetV2 \cite{mobilenetv2}, adopt DSCs. 

\subsection{Accelerating CNN Tensor Products}\label{accCNN-TP}
Modern CNNs employ a large number of input and kernel tensors across all their constituent standard convolutional, depthwise seperable convolutional and fully-connected layers. These CNN tensors are often irregular in size. For example, ResNet50\cite{resnet50} (EfficientNetB7\cite{efficientnet}) CNN employs kernel tensors of at least 12 (26) different sizes (see the sizes in Table \ref{dsc_tensor_info}). Accelerating the inference of such CNNs having mixed-sized tensors requires efficient hardware implementations of the Products (a.k.a. General Matrix Multiplications (GEMMs)) of their input and kernel (weight) tensors. To make the hardware implementations of such CNN Tensor Products feasible, the involved input and kernel tensors are typically decomposed into vectors (1D tensors). These vectors are referred to as Decomposed Input Vectors (DIVs) and Decomposed Kernel Vectors (DKVs), henceforth. Decomposing into DIVs and DKVs in turn transforms the Tensor Products into decomposed vector dot products (VDPs), which can be efficiently performed on VDP hardware accelerators \cite{17}. When implemented on hardware, these decomposed VDPs produce partial results (\textit{psums}), which can then be post-processed to achieve the final Tensor Product results using the \textit{psum} reduction networks employed in the hardware accelerators \cite{17}.

\begin{figure}[h]
  \centering
  \includegraphics[scale = 0.35]{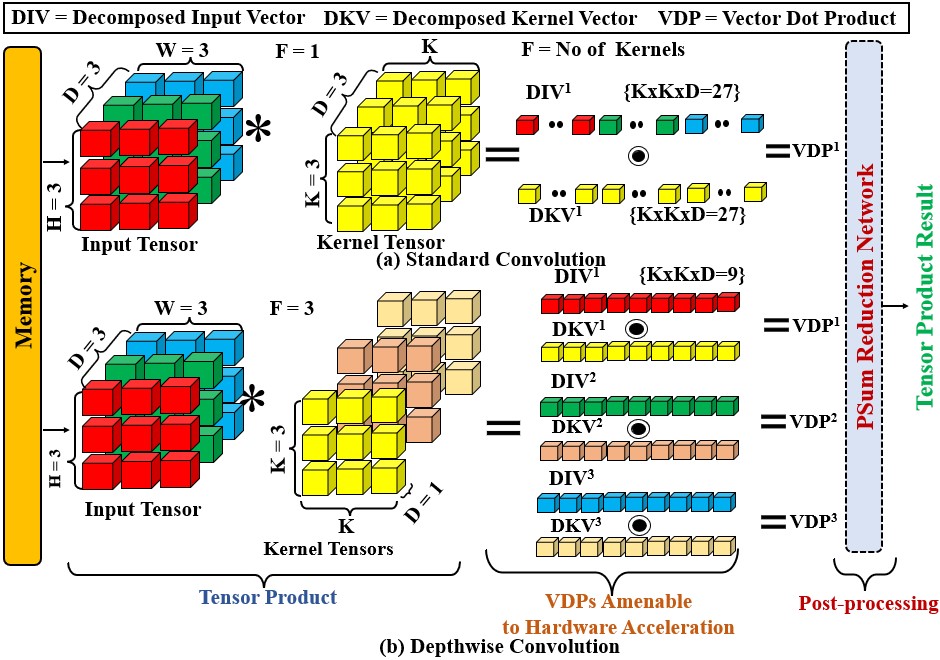}
  \caption{Illustration of decomposition of CNN tensors for tensor product acceleration, for (a) standard convolution (SC), and (b) depthwise convolution (DC).}
  \label{fig2}
\end{figure}

To conceive this process of Tensor Product acceleration, consider Fig. \ref{fig2}, which illustrates tensor products between the input and kernel tensors, for a SC (Fig. \ref{fig2}(a)) and a DC (Fig. \ref{fig2}(b)). In Fig \ref{fig2}(a), the tensor product between an input tensor of dimension (H, W, D) = (3, 3, 3) and a single kernel tensor (F = 1) of dimension (K, K, D) = (3, 3, 3) is illustrated. To make this tensor product amenable to hardware acceleration, these input and kernel tensors are respectively flattened into a DIV ($DIV^1$) and a DKV ($DKV^1$) of 27 points each (corresponding to K$\times$K$\times$D = 3$\times$3$\times$3). Consequently, the tensor product is converted into a VDP operation between the $DIV^1$ and $DKV^1$, which produces $VDP^1$ as the result. Producing this $VDP^1$ can be simplified into 27 point-wise multiplication operations to produce 27 point-wise products, which are then summed together using 27 accumulation operations. This makes producing $VDP^1$ amenable to acceleration on any computing hardware that can support multiple parallel multiplications and accumulations (e.g., \cite{analogelectronic}, \cite{digitalelectronic}, \cite{inmemory}, \cite{fpga}, \cite{opticalanalog2020}). If the number of supported, in-parallel multiplications and accumulations in the employed accelerator hardware is less than the size of the VDP operation (27 in our example), the result $VDP^1$ has to be decomposed into multiple \textit{psum} results. These \textit{psum} results are then summed together using the \textit{psum} reduction network (Fig. \ref{fig2}), to produce the final tensor product result, at the cost of extra latency. Note that for the example illustrated in Fig. \ref{fig2}, we assume that the accelerator hardware size matches with the VDP operation size.

In contrast, as discussed earlier, performing a DC operation on the same input tensor of dimension (H, W, D)=(3, 3, 3) requires 3 channel-wise kernel tensors (F=3) of dimension (K, K, D) = (3, 3, 1) each. The need to use 3 channel-wise kernel tensors necessitates that a total of 3 tensor products are obtained. For that, as shown in Fig. \ref{fig2}(b), the 3 channels of the input tensor are flattened into 3 DIVs ($DIV^1$, $DIV^2$, $DIV^3$), and the 3 channel-wise kernel tensors are flattened into 3 DKVs ($DKV^1, DKV^2, DKV^3$), of 9 points each (corresponding to K$\times$K$\times$D = 3$\times$3$\times$1). Performing VDP operations between respective DIVs and DKVs produces 3 VDP results ($VDP^1$, $VDP^2$, $VDP^3$). Thus, compared to an SC operation, a DC operation renders smaller sized DIVs and DKVs. This in turn reduces the parallelism requirement per VDP result in the employed accelerator hardware, because implementing the VDP operations between the reduced sized DIVs and DKVs requires the hardware support for a less number of in-parallel multiplications and accumulations per VDP result (only 9 in our example of DC). If the employed accelerator hardware supports more parallelism than necessary for implementing a DC operation, it can lead to lower hardware utilization efficiency. Similar to the SC and DC operations, it is also common to convert a PC operation into multiple VDP operations to make it amenable to hardware acceleration. \textit{In summary}, to make processing of CNN tensor products amenable to hardware-based acceleration, this process of tensor flattening and decomposition remains consistent for the input and kernel tensors of arbitrary sizes (for sizes other than considered in Fig. \ref{fig2} as well), across all types of convolutional and fully-connected CNN layers.

\subsection{Related Work on Photonic CNN Accelerators}
To accelerate machine learning tasks with low latency and low energy consumption, prior works have proposed various accelerators based on Photonic Integrated Circuits (PICs) (e.g., \cite{11,15,14,13,17,mzi2018,mzicomplex2021,sairam2021}). Among these, the CNN accelerators employ PIC-based TPCs to perform CNN tensor products. Some accelerators implement digital TPCs (e.g., \cite{28,albireo}), whereas some others employ analog TPCs (e.g., \cite{17,10,11,tait2017}). Different TPC implementations employ MRRs (e.g., \cite{14,13,17,tait2020photonic,feldmannMRR2021parallel}) or MZIs (e.g., \cite{16,mzi2018,mzicomplex2021}) or both (e.g., \cite{28}). The analog TPCs can be further classified as incoherent (e.g., \cite{17,14,11}) or coherent (e.g., \cite{16,coherentrayhamerly2019,coherentZhao2019,coherentnature21,coherentZhao18,coherent2018,chorentzhou2021large}). To set and update the values of the individual input and kernel tensors used for tensor products, the incoherent TPCs utilize the analog optical signal power, whereas the coherent TPCs utilize the electrical field amplitude and phase. The coherent TPCs achieve low inference latency, but they suffer from control complexity, high area overhead, low scalability, low flexibility, high encoding noise, and phase error accumulation issues \cite{29}. In contrast, the incoherent TPCs based accelerators achieve better scalability and lower footprint, because they use MRR-based compact PICs \cite{17}, unlike the coherent TPCs that use MZI-based bulky PICs.

Various state-of-the-art photonic CNN accelerators are well discussed in a survey paper \cite{30}. Because of the inherent advantages of MRR-enabled incoherent TPCs, there is impetus to design more energy-efficient and scalable CNN accelerators employing MRR-enabled incoherent TPCs. However, the CNN accelerators from prior works have mainly focused on designing their constituent TPCs only for the standard convolutional layers of CNNs. Prior works have paid very little attention to accelerate the processing of the depthwise separable convolutional layers of CNNs. In this paper, we contribute towards making the MRR-enabled incoherent TPCs more efficient by enabling them to dynamically adapt to process the standard convolutional and depthwise separable convolutional layers of CNNs.   

\begin{figure}[h!]
  \centering
  \includegraphics[scale=0.55]{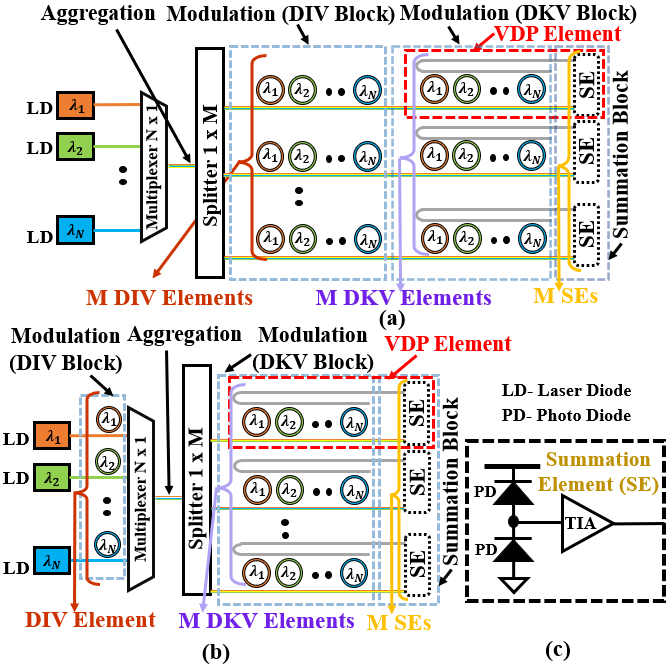}
  \caption{Illustration of common TPC organizations. (a) AMM organization, (b) MAM organization, and (c) Summation Element.}
  \label{fig3ab}
\end{figure}

\section{Classification and Scalability Analysis of TPC Organizations}
\subsection{Classification}\label{sec3}
Most of the photonic MRR-enabled analog, incoherent CNN accelerators from prior works employ multiple analog TPCs that work in parallel. Typically, an analog TPC implements the decomposed VDP operations of a tensor product (Fig. \ref{fig2}; Section \ref{accCNN-TP}). For that, the TPC typically employs a total of five blocks (Fig. \ref{fig3ab}(a)): \textit{(i)} a laser block that employs \textit{N} Laser Diodes (LDs) to generate \textit{N} optical wavelength channels; \textit{(ii)} an aggregation block that aggregates the generated optical wavelength channels into a single photonic waveguide through DWDM (using an \textit{N}$\times$1 multiplexer) and then splits the optical power of these \textit{N} wavelength channels equally into \textit{M} separate waveguides (using a 1$\times$\textit{M} splitter); \textit{(iii)} a modulation block, also referred to as DIV block, that employs \textit{M} arrays of MRRs (one array per waveguide, with each array having \textit{N} MRRs; each array referred to as DIV element) to imprint \textit{M} DIVs of \textit{N} points each onto the \textit{N}$\times$\textit{M} wavelength channels by modulating the analog power amplitudes of the wavelength channels; \textit{(iv)} another modulation block, referred to as DKV block, that employs another \textit{M} arrays of MRRs (one array per waveguide, with each array having \textit{N} MRRs; each array referred to as DKV element) to further modulate the \textit{N}$\times$\textit{M} wavelength channels with DKVs, so that the analog power amplitudes of the individual wavelength channels then represent the point-wise products of the utilized DKVs and DIVs; and \textit{(v)} a Summation Block (SB) that employs a total of \textit{M} Summation Elements (SEs), with each SE having two balanced Photodiodes (PDs) upon which the point-wise-product-modulated \textit{N} wavelength channels are incident to produce the output current that is proportional to the VDP result (i.e., the sum of the \textit{N} input point-wise products). The laser block and SB are typically positioned at the two ends of the TPC, with the aggregation, modulation (DIV), and modulation (DKV) blocks placed in between them. 

\begin{table*}[]
\centering
\caption{List of abbreviations and their full forms used in this paper. Definition and values of various parameters (obtained from \cite{19}) used in Eq. \ref{eq1}, Eq. \ref{eq1a}, and Eq. \ref{eq2} for the scalability analysis of AMM and MAM TPCs.}
\label{abbrevations}
\begin{tabular}{|c|c||c|c|c|}
\hline
{ \textbf{Abbreviations}} & { \textbf{Full   form}}                                                                             & { \textbf{Parameter}}             & { \textbf{Definition}}                                                                                                         & { \textbf{Value}}   \\ \hline
{ TPC}           & { Tensor Processing Core}                                                                  & { $P_{Laser}$}                & { Laser Power Intensity}                                                                                               & { 10 dBm}  \\ \hline
{ DSC}           & { Depthwise Separable Convolution}                                                         & { R}                     & { PD Responsivity}                                                                                                     & { 1.2 A/W} \\ \hline
{ SC}            & { Standard Convolution}                                                                    & { $R_L$}                    & { Load Resistance}                                                                                                     & { 50 $\Omega$}      \\ \hline
{ PC}            & { Pointwise Convolution}                                                                   & { $I_d$}                    & { Dark Current}                                                                                                        & { 35 nA}   \\ \hline
{ DC}            & { Depthwise Convolution}                                                                   & { T}                     & { Absolute Temperature}                                                                                                & { 300 K}   \\ \hline
{ DKV}           & { Decomposed Kernel Vector}                                                                & { BR}                    & { Bit Rate}                                                                                                            & { 10 GS/s} \\ \hline
{ DIV}           & { Decomposed Input Vector}                                                                 & { RIN}                   & { Relative Intensity Noise}                                                                                            & { -140 dB/Hz}   \\ \hline
{ VDP}           & { Vector Dot Product}                                                                      & { $\eta_{WPE}$}              & { Wall Plug Efficiency}                                                                                                & { 0.1}     \\ \hline
{ S}             & { Size of DKV}                                                                             & { $IL_{SMF}$(dB)}           & { Single Mode Fiber Insertion Loss}                                                                                    & { 0}       \\ \hline
{ $DKV_S$}          & { DKV of size S}                                                                           & { $IL_{EC}$(dB)}            & { Fiber to Chip Coupling Insertion Loss}                                                                               & { 1.6}     \\ \hline
{ $F_S$}            & { Set of DKVs with size S}                                                                 & { $IL_{WG}$(dB/mm)}         & { Silicon Waveguide Insertion Loss}                                                                                    & { 0.3}     \\ \hline
{ \textit{BR}}          & { Bit rate}                                                                             & { $EL_{Splitter}$(dB)}      & { Splitter Insertion Loss}                                                                                             & { 0.01}    \\ \hline
{ VDPE}          & { Vector Dot Product Element}                                                              & { $IL_{MRM}$(dB)}           & { Microring Modulator (MRM) Insertion Loss}                                                                            & { 4}       \\ \hline
{ \textit{N}}             & { Size of VDPE}                                                                            & { $OBL_{MRM}$(dB)}          & { Out of Band Loss MRM}                                                                                                & { 0.01}    \\ \hline
{ \textit{M}}             & { Number of VDPEs per TPC Unit}                                                            & { $IL_{MRR}$(dB)}           & { Microring Resonator (MRR) Insertion Loss}                                                                             & { 0.01}    \\ \hline
{ SE}            & { Summation Element}                                                                       & { $IL_{penalty}$(dB) (MAM)} & { Network Penalty}                                                                                                     & { 4.8}     \\ \hline
{ CS}            & { Comb Switch}                                                                             & { $IL_{penalty}$(dB) (AMM)} & { Network Penalty}                                                                                                     & { 5.8}     \\ \hline
{ \textit{y}}             & { Number of Comb Switches}                                                                 & { $d_{MRR}$}                & { Gap between two adjacent MRRs}                                                                                       & { 20 $\mu$m}   \\ \hline
{ \textit{x}}             & { \begin{tabular}[c]{@{}c@{}}Re-aggregating or Filtering \\ wavelength count\end{tabular}} & { $d_{element}$ (MAM)}      & { }                                                                                                                    & { 0 $\mu$m}    \\ \cline{1-3} \cline{5-5} 
{ \textit{L}}             & { Set of re-aggregated wavelengths}                                                        & { $d_{element}$ (AMM)}      & \multirow{-2}{*}{{ \begin{tabular}[c]{@{}c@{}}Thermal isolation spacing \\ between DIV and DKV elements\end{tabular}}} & { 100 $\mu$m}   \\ \hline
\end{tabular}
\end{table*}

Based on the order in which these intermediate blocks (aggregation, modulation (DIV), modulation (DKV) blocks) are positioned between the laser block and SB, we classify the MRR-based TPC organizations from prior work as MAM (Modulation, Aggregation, Modulation) \cite{17} or AMM (Aggregation, Modulation, Modulation) \cite{15}. Fig. \ref{fig3ab} illustrates MAM and AMM TPC organizations. From Fig. \ref{fig3ab}(a), the AMM TPC organization positions the aggregation block first, and then the DIV block followed by the DKV block. In contrast, the MAM TPC in Fig. \ref{fig3ab}(b) positions the DIV block first, and then positions the aggregation block followed by the DKV block. Note that the MAM-styled DIV block is structurally different from the AMM-styled DIV block. The MAM-styled DIV block employs only one MRR per waveguide, and as a result, it can imprint only 1 DIV of \textit{N} points onto the \textit{N} wavelength channels. This 1 DIV is shared among all DKVs in the MAM TPC, whereas each DKV can have a different DIV corresponding to it in the AMM TPC.

Moreover, note that each DKV element in both MAM and AMM TPCs have two waveguides (shown but not labelled in the figures). First, the drop waveguide, coupling with the MRRs at the top. Second, the through waveguide, coupling with the MRRs at the bottom. The wavelengths that carry negatively-signed pointwise products have more guided optical power in the drop waveguide, whereas the wavelengths that carry positively-signed pointwise products have more guided optical power in the through waveguide. The drop waveguide makes its guided power incident upon the top-side PD in the corresponding SE, whereas the through waveguide makes its guided power incident upon the bottom-side PD. This enables a signed accumulation of the pointwise products carried by the guided wavelengths, because the top-side and bottom-side PDs in the SE are balanced \cite{15}. In both the AMM and MAM TPC organizations, we refer to the combination of a DKV element  and the corresponding SE as VDP element (VDPE). A VDPE size (i.e., \textit{N}) should match the DKV size for efficient, low-overhead implementation of the VDP operation.

\subsection{Scalability Analysis}\label{sec3b}
Prior works \cite{19} and \cite{albireo} have shown that the scalability of photonic accelerator architectures, in terms of the achievable VDPE size \textit{N}, decreases with the increase in the required bit precision. However, these prior works lack in two ways. First, they do not capture the impact of the utilized bit rate on the inter-dependence between \textit{N} and bit precision. Second, they do not provide the scalability analysis for AMM TPC architectures (they only analyze MAM TPCs). To address these shortcomings, we extend the methodology from \cite{19} to perform the scalability analysis of the AMM and MAM TPCs to capture the inter-dependence of VDPE size \textit{N}, number of VDPEs per TPC \textit{M}, bit precision, and bit rate. From \cite{19}, it is known that the bit precision of an MRR-based VDPE depends on the output photodetector sensitivity ($P_{PD-opt}$) and bit rate (BR) \cite{19}. In this paper, we evaluate the required $P_{PD-opt}$ for various bit precision values and BRs by solving Eq. \ref{eq1} \cite{19}. We sweep for BRs of 1 Gbps, 3 Gbps, 5 Gbps, and 10 Gbps, and sweep for bit precision values of 1-bit to 8-bit, and consider \textit{M = N} for our analysis. The value of \textit{N} is obtained by solving Eq. \ref{eq2} \cite{19}, along with $P_{PD-opt}$ for a given bit precision and BR obtained from Eq. \ref{eq1}  and Eq. \ref{eq1a} \cite{19}.  

\begin{equation}
       n_{i/p} = \frac{1}{6.02}\Bigg[20log_{10}(\frac{R\times P_{PD-opt}}{\beta\sqrt{\frac{BR}{\sqrt{2}}}}-1.76\Bigg]
       \label{eq1}
\end{equation}

\begin{equation}
    \beta = \sqrt{2q(RP_{PD-opt}+I_d)+\frac{4kT}{R_L}+R^2P_{PD-opt}^2RIN}
    \label{eq1a}
\end{equation}

\begin{equation}
   \label{eq2}
\begin{split}
   P_{Laser} = \frac{10^{\frac{\eta_{WG}(dB)[N(d_{MRR})+d_{element}]}{10}}M}{\eta_{SMF}\eta_{EC}IL_{i/p-MRM}}
    \times\frac{P_{PD-opt}}{\eta_{WPE}IL_{MRR}}
      \\\times\frac{1}{(OBL_{MRM})^{N-1}(EL_{splitter})^{log_{2}M}} 
    \\\times \frac{1}{(OBL_{MRR})^{N-1}(IL_{penalty})}
\end{split}
\end{equation}

Table \ref{abbrevations} reports the definitions of the parameters and their values from Eq. \ref{eq1}, Eq. \ref{eq1a}, and Eq. \ref{eq2}, as used for our analysis. In Table \ref{abbrevations}, $P_{penalty}$ represents the impairments due to extinction ratio, crosstalk, inter-symbol interference (ISI), and laser relative intensity noise (RIN). We perform this analysis for both AMM and MAM TPC architectures. As discussed in section \ref{sec3}, AMM and MAM TPCs differ in the placement of the DKV and DIV elements. In MAM TPCs, all DKV elements share a single DIV element, whereas in AMM TPCs all DKV elements have their individual DIV elements. Therefore, to avoid thermal crosstalk in AMM TPCs, the DKV elements need to be placed sufficiently farther from their respective DIV elements \cite{keren2020}, which in turn increases the required $d_{element}$ for AMM TPCs (Table \ref{abbrevations}), thereby increasing the optical lengths of the waveguides in AMM TPCs. Due to the longer waveguides, the $IL_{penalty}$ increases for AMM TPCs compared to MAM TPCs (Table \ref{abbrevations}). This in turn affects the achievable VDPE size \textit{N}, BR, and bit precision for AMM TPCs, as confirmed by our below-presented analysis results.

 \begin{figure} [H]
        \centering
        
        \includegraphics[scale=0.85]{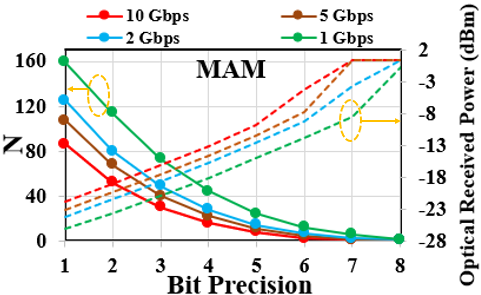}
        \caption{Supported VDPE size \textit{N} and the optical received power (dBm) for bit precision =\{1, 2, 3, 4, 5, 6, 7, 8\}bits at bit rates (BRs) = \{1, 3, 5, 10\}Gbps, for MAM TPCs.}
       \label{scalabilityMAM}
    \end{figure}
   
     \begin{figure} [H]
        \centering
         
        \includegraphics[scale=0.85]{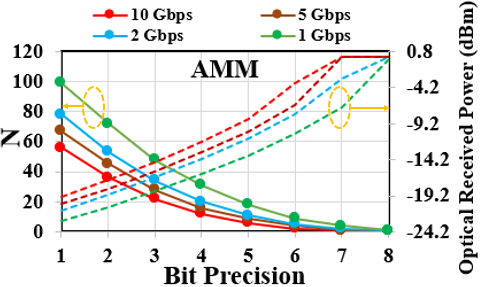}
        \caption{Supported VDPE size \textit{N} and the optical received power (dBm) for bit precision =\{1, 2, 3, 4, 5, 6, 7, 8\}bits at bit rates (BRs) = \{1, 3, 5, 10\}Gbps, for AMM TPCs.}
      \label{scalabilityAMM}
    \end{figure}

Fig. \ref{scalabilityMAM} and Fig. \ref{scalabilityAMM} show the inter-dependence of VDPE size \textit{N}, bit precision, and BR for MAM and AMM TPCs respectively. From Fig. \ref{scalabilityMAM}, for MAM TPCs, the maximum VDPE size \textit{N} significantly decreases from \textit{N = 159} for 1-bit precision to \textit{N = 1} for 8-bit precision. Similarly, from Fig. \ref{scalabilityAMM}, for AMM TPCs, the supported \textit{N} drops from \textit{N = 99} for 1-bit precision to \textit{N = 1} for 8-bit precision. This trend of decreasing \textit{N} with increasing bit precision is in line with the similar trend observed in recent work Albireo \cite{albireo}, although note that the accelerator architecture of Albireo is different from our considered AMM and MAM architectures. Moreover, we can also observe from Fig. \ref{scalabilityMAM} and Fig. \ref{scalabilityAMM} that the supported \textit{N} also varies with BR. With the increase in BR, the supported \textit{N} decreases in both MAM and AMM TPCs. For instance, MAM TPCs' supported \textit{N} for 4-bit precision drops from \textit{N = 44} at 1 Gbps to \textit{N = 16} at 10 Gbps. In case of AMM TPCs, at 4-bit precision, the supported \textit{N} drops from \textit{N = 31} at 1 Gbps to \textit{N = 12} at 10 Gbps. For given bit precision and BR values, AMM TPCs support lower \textit{N} compared to MAM TPCs. This is because, as discussed earlier, AMM TPCs incur higher $IL_{penalty}$. From our analysis, it can be inferred that AMM and MAM TPCs cannot support any \textit{N} for 8-bit precision; therefore, we advocate that AMM and MAM TPCs should be utilized to achieve 4-bit precision at the highest, to select such a power-of-two precision value below the unattainable 8-bit precision that can support a tangible \textit{N} value. Our obtained values of \textit{N} for different BRs are given in Table \ref{Nvalues}.     
 
\medskip

\begin{table}[H]
\centering

\caption{VDPE size (\textit{N}) at 4-bit precision across various BRs for different TPC architectures.}
\label{Nvalues}
\begin{tabular}{|c|llll|}
\hline
\multicolumn{1}{|c|}{\multirow{2}{*}{\textbf{TPC Architectures}}} & \multicolumn{4}{c|}{\textbf{BR   (Gbps)}}                                                                         \\ \cline{2-5} 
\multicolumn{1}{|l|}{}                  & \multicolumn{1}{l|}{\textbf{1}} & \multicolumn{1}{l|}{\textbf{3}} & \multicolumn{1}{l|}{\textbf{5}} & \textbf{10} \\ \hline \hline
\textbf{RMAM}                           & \multicolumn{1}{l|}{43}         & \multicolumn{1}{l|}{27}         & \multicolumn{1}{l|}{22}         & 16          \\ \hline
\textbf{RAMM}                           & \multicolumn{1}{l|}{31}         & \multicolumn{1}{l|}{20}         & \multicolumn{1}{l|}{16}         & 12          \\ \hline
\textbf{MAM (HOLYLIGHT  \cite{17})}     & \multicolumn{1}{l|}{44}         & \multicolumn{1}{l|}{28}         & \multicolumn{1}{l|}{22}         & 16          \\ \hline
\textbf{AMM (DEAPCNN   \cite{15})}       & \multicolumn{1}{l|}{31}         & \multicolumn{1}{l|}{20}         & \multicolumn{1}{l|}{16}         & 12          \\ \hline
\end{tabular}
\end{table}



\section{Need for Reconfigurability in TPCs}\label{need_for_reconfigurability}

A feasible acceleration of CNN tensor products on MRR-enabled incoherent, analog TPCs mandates that a CNN kernel tensor of shape (K, K, D) is decomposed/flattened into a 1D DKV of size \textit{S} = K$\times$K$\times$D. From Fig. \ref{fig2}, the corresponding DIV should also be of size \textit{S}. In modern CNNs, the value \textit{S} corresponding to various kernel tensors vary drastically. Table \ref{dsc_tensor_info} provides kernel tensor shapes and corresponding DKV sizes (\textit{S}) for EfficientNetB7 \cite{efficientnet}. We selected EfficientNetB7 as an example of CNNs with a large number of DCs, PCs, and SCs. From Table \ref{dsc_tensor_info}, DCs have a small set of DKV sizes (\textit{S} $\in$ \{9, 25\}). In contrast, in Table \ref{dsc_tensor_info}, PCs have a wide range of \textit{S} values, from as small as 8 to as large as 3840. As discussed earlier (Section II-B), for the processing of PCs and DCs using TPCs to produce the final tensor products, their corresponding DKVs are mapped onto the constituent VDPEs of the TPCs so that the tensor products are converted into VDP operations. Therefore, depending on how the size \textit{S} of a DKV compares with the size \textit{N} of a VDPE, the following three scenarios arise for the mapping of the DKV onto the VDPE to produce the final VDP result:



\begin{itemize}

    \item \textbf{Scenario 1, \textit{S}=\textit{N}}: For this case, all the DKV points have one-to-one mapping with all the MRRs of the VDPE, and there are no idle MRRs in the VDPE. The VDP result will be the final tensor product result. 
    
    \item \textbf{Scenario 2, \textit{S}$>$\textit{N}}: In this case, a single VDPE cannot produce the final tensor product result, as it cannot process the entire DKV. Therefore, the DKV needs to be further decomposed into a total \textit{P} partial DKVs requiring a total of \textit{P} = \textit{Ceil}(\textit{S}/\textit{N}) VDPEs to produce the final tensor product result. All P VDPEs produce partial VDP results called \textit{psum}s, and during post-processing, a reduction network accumulates these \textit{psum}s to generate the final tensor product result. Although this accumulation of \textit{psum}s incurs additional latency, which can be efficiently hidden by employing a pipelined design of the \textit{psum} reduction network with non-blocking bandwidth \cite{maeri_asplos2018}. One drawback, however, is that if \textit{S} is not an integer multiple of \textit{N}, then this scenario leads to some unutilized MRRs in the VDPEs across \textit{P} partial VDP operations.
    
    \item \textbf{Scenario 3, \textit{S}$<$\textit{N}}: In this case, the result of the single VDP operation provides the final tensor product result, but some MRRs in the VDPE remain unutilized. The count of unutilized MRRs depends on the size difference between the DKV and VDPE (i.e., between \textit{S} and \textit{N}). 
\end{itemize}

For the last two scenarios (for which \textit{S}$\neq$\textit{N}), the unutilized MRRs cause underutilization in the VDPEs. This hampers the performance and efficiency of processing modern CNNs with mixed-sized tensors. This is because the unutilized MRRs incur area and static power overheads while also idling away the opportunity for increasing the processing throughput. Therefore, how well \textit{N} matches with \textit{S} plays a crucial role in determining the performance and efficiency of a TPC. To that end, we reason that dynamic flexibility in the supported value of \textit{N} in the VDPEs is required to efficiently support processing of various sizes of DCs, PCs, and SCs of modern CNNs.
\break

\begin{table}[H]
\centering 
\caption{Kernel tensor shapes (K, K, D), the total number of such kernels (F) and corresponding DKV sizes (\textit{S}) for EfficientNet\_B7\cite{efficientnet}, as an example CNN with a large number of DSCs. (FC=Fully Connected Layer and other abbreviations are defined in Table \ref{abbrevations}). The K, D, F values were extracted from Keras Applications \cite{chollet2015keras}.}
\label{dsc_tensor_info}
\begin{tabular}{|c|c|c|c|c|}
\hline
\textbf{Model}                     & \textbf{Convolution} & \textbf{\begin{tabular}[c]{@{}c@{}}Tensor Shape\\ (K, K, D)\end{tabular}} & \textbf{F} & \textbf{S} \\ \hline
\multirow{26}{*}{EfficientNet\_B7} & DC                   & (3, 3, 1)                                                                 & 25024      & 9          \\ \cline{2-5} 
                                   & DC                   & (5, 5, 1)                                                                 & 45216      & 25         \\ \cline{2-5} 
                                   & PC                   & (1, 1, 8)                                                                 & 288        & 8          \\ \cline{2-5} 
                                   & PC                   & (1, 1, 12)                                                                & 2016       & 12         \\ \cline{2-5} 
                                   & PC                   & (1, 1, 16)                                                                & 64         & 16         \\ \cline{2-5} 
                                   & PC                   & (1, 1, 20)                                                                & 3360       & 20         \\ \cline{2-5} 
                                   & PC                   & (1, 1, 32)                                                                & 312        & 32         \\ \cline{2-5} 
                                   & PC                   & (1, 1, 40)                                                                & 9600       & 40         \\ \cline{2-5} 
                                   & PC                   & (1, 1, 48)                                                                & 2016       & 48         \\ \cline{2-5} 
                                   & PC                   & (1, 1, 56)                                                                & 13440      & 56         \\ \cline{2-5} 
                                   & PC                   & (1, 1, 64)                                                                & 48         & 64         \\ \cline{2-5} 
                                   & PC                   & (1, 1, 80)                                                                & 3360       & 80         \\ \cline{2-5} 
                                   & PC                   & (1, 1, 96)                                                                & 29952      & 96         \\ \cline{2-5} 
                                   & PC                   & (1, 1, 160)                                                               & 21120      & 160        \\ \cline{2-5} 
                                   & PC                   & (1, 1, 192)                                                               & 56         & 192        \\ \cline{2-5} 
                                   & PC                   & (1, 1, 224)                                                               & 13440      & 224        \\ \cline{2-5} 
                                   & PC                   & (1, 1, 288)                                                               & 452        & 288        \\ \cline{2-5} 
                                   & PC                   & (1, 1, 384)                                                               & 29952      & 384        \\ \cline{2-5} 
                                   & PC                   & (1, 1, 480)                                                               & 780        & 480        \\ \cline{2-5} 
                                   & PC                   & (1, 1, 640)                                                               & 14080      & 640        \\ \cline{2-5} 
                                   & PC                   & (1, 1, 960)                                                               & 2064       & 960        \\ \cline{2-5} 
                                   & PC                   & (1, 1, 1344)                                                              & 2960       & 1344       \\ \cline{2-5} 
                                   & PC                   & (1, 1, 2304)                                                              & 6496       & 2304       \\ \cline{2-5} 
                                   & PC                   & (1, 1, 3840)                                                              & 2400       & 3840       \\ \cline{2-5} 
                                   & SC                   & (3, 3, 3)                                                                 & 64         & 27         \\ \cline{2-5} 
                                   & FC                   & (2560, 1, 1)                                                              & 1          & 2560       \\ \hline
\end{tabular}
\end{table}

\subsection{Perils of Fixed VDPE Size (N) in TPCs from Prior Work}
\begin{figure}[h]
  \centering
  \includegraphics[width=\linewidth]{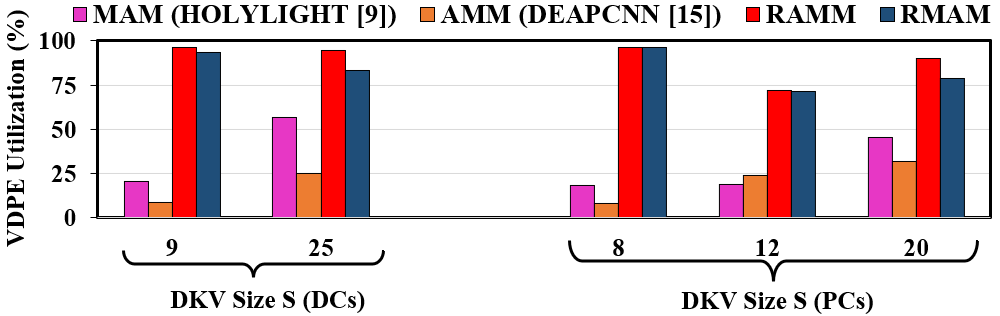}
  \caption{VDPE utilization \% (utilized VDPE area/total area) for MAM (HOLYLIGHT\cite{17},N=44), AMM (DEAPCNN\cite{15},N=31), RAMM (N=31), and RMAM (N=43) at BR=1Gbps and 4-bit precision for various DKV sizes corresponding to DCs and PCs.}
  \label{vdpe_utilization}
\end{figure}

To validate our reasoning presented just before this subsection, we evaluated the hardware utilization values for various TPC architectures (with fixed \textit{N}) from prior works. For example, MAM (HOLYLIGHT \cite{17}), AMM (DEAPCNN \cite{15}) have \textit{N}=43 and \textit{N}=31,  respectively. For these TPCs, Fig. \ref{vdpe_utilization} shows hardware (MRR) utilization per-VDPE in terms of the ratio (in \%) of the utilized VDPE area over the total (utilized+idle) area. The figure also shows the utilization for our proposed Reconfigurable MAM (RMAM) and Reconfigurable AMM (RAMM) TPCs. But we introduce and discuss our RAMM and RMAM TPCs in Section \ref{reconfig}, so please look past their excellent utilization results for now. From Fig. \ref{vdpe_utilization}, MAM (HOLYLIGHT \cite{17}), and AMM (DEAPCNN \cite{15}) yield significantly low per-VDPE utilization (as low as 8\%). This is because of the huge mismatch between \textit{N} and \textit{S} (i.e., \textit{S}$<$\textit{N}) while processing DCs and PCs. Such low VDPE utilization can hamper performance and efficiency of the TPCs, as discussed earlier. 

These results motivate the need to design a VDPE that can dynamically adapt to various DKV sizes. Therefore, we invented a novel reconfigurable VDPE design, which is described in the next section.

\section{MRR-based Reconfigurable TPC Architectures}\label{reconfig}
In this section, we present a novel reconfigurable VDPE which serves as the backbone of our proposed MRR-based reconfigurable TPC architectures. This reconfigurable VDPE adds the following two desirable attributes to the AMM and MAM types of TPCs. \textit{First}, it introduces the flexibility to the TPCs so that the processing of the DKVs of various sizes can be efficiently mapped onto them, regardless of the fixed VDPE size \textit{N} for the TPCs. \textit{Second}, it introduces opportunities for increasing the processing parallelism and MRR utilization efficiency. This reconfigurable VDPE can directly replace the VDPEs of the MAM and AMM TPCs (Fig. \ref{fig3ab}), to convert them into reconfigurable MAM (we refer to it as RMAM, henceforth) and reconfigurable AMM (we refer to it as RAMM, henceforth) TPCs. The structure and operation of our invented reconfigurable VDPE are discussed in the following subsections.

\subsection{Reconfigurable VDPE: Structure and Layout}
\begin{figure}[h]
  \centering
  \includegraphics[scale= 0.7]{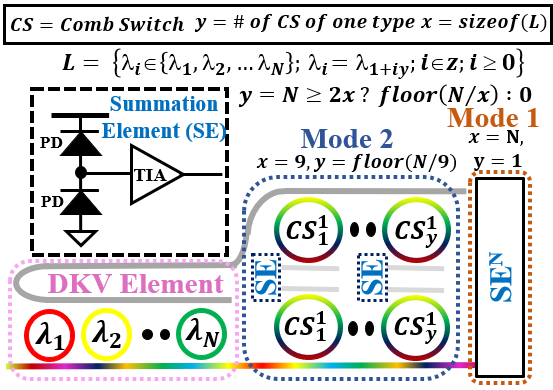}
  \caption{Schematic of our invented reconfigurable VDPE, employing a DKV element and one group of comb switch (CS) pairs corresponding to the reconfiguration Mode 2.}
  \label{fig4}
\end{figure}

Fig. \ref{fig4} illustrates our proposed reconfigurable VDPE. It consists of a DKV element, which is an array of \textit{N} modulation MRRs that can imprint \textit{N} pointwise products onto the incoming \textit{N} wavelength channels, similar to the DKV elements of the MAM and AMM TPCs from Fig. \ref{fig3ab}. This DKV element is followed by a group of a total of \textit{y} pairs of MRR comb switches (CSs). These \textit{y} CS pairs can re-aggregate the incoming \textit{N} pointwise-product-modulated wavelength channels (incoming from the DKV element) into a total of \textit{y} distinct sets, with each set \textit{L} (Fig. \ref{fig4}) having a total of \textit{x} distinct wavelength channels. In other words, each CS pair filters a comb (set \textit{L}; Fig. \ref{fig4}) of \textit{x} distinct wavelength channels from the incoming \textit{N} wavelength channels. Each CS pair is able to do this because of its innate spectral response that allows it to be in resonance with \textit{x} distinct wavelengths simultaneously (more on the design of CSs in Section V.C). Each CS pair then sends its corresponding \textit{x} wavelength channels to its dedicated summation element (SE), which performs a signed accumulation of the data carried on the \textit{x} wavelength channels to produce a VDP result. To enable the signed accumulation, similar to the SEs of the AMM and MAM TPCs (Fig. \ref{fig3ab}), the SEs corresponding to the CS pairs also employ balanced PDs. Thus, the group of \textit{y} CS pairs per reconfigurable VDPE enables \textit{y} VDP results of size \textit{x} each to be produced in parallel. 

Note that here \textit{x}, \textit{y}, and \textit{N} are integers, and their relation is given by this equation: \textit{\textbf{y = N$>$2x $?$ floor(N/x) : 0}}. Henceforth, we refer to \textit{x} as re-aggregation size. Also note that for the group of CS pairs to produce \textit{y} in-parallel VDP results, each CS pair in the group has to be switched ON by electro-optically tuning its spectral resonance passbands to align with its corresponding set of \textit{x} wavelength channels \cite{combswitch}. In contrast, it is also possible to switch OFF a CS pair by tuning its resonance passbands out of alignment with the corresponding \textit{x} wavelength channels. When the CS pairs are switched OFF, all of the \textit{N} incoming wavelength channels are allowed to pass by the CS pairs to be eventually accumulated at the summation element SE$^N$. Thus, depending on whether the CS pairs are ON or OFF, the reconfigurable VDPE can operate in two different modes. These modes are further explained in the next section.

\subsection{Reconfigurable VDPE: Operation}

\begin{figure}[h]
  \centering
  \includegraphics[width=\linewidth]{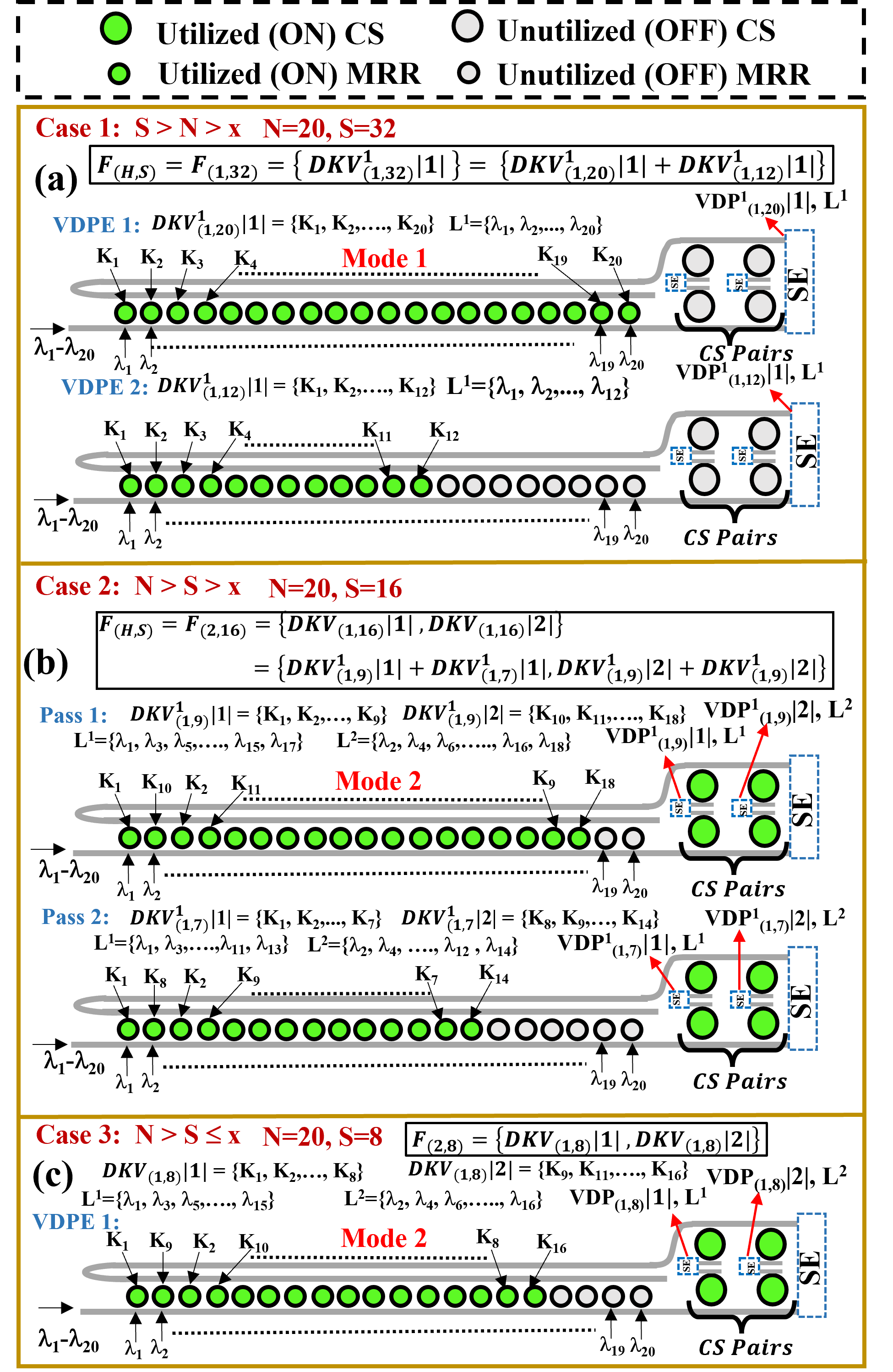}
  \caption{Example operation of our reconfigurable VDPE for various cases depending on the \textit{S} and \textit{N} values, for \textit{x=9}. Here, ON and OFF CS pairs, respectively, represent Mode 2 and Mode 1 of operation.}
  \label{rvdpeoperation}
\end{figure}

The reconfigurable VDPEs of our RMAM and RAMM TPC architectures support two operational modes, i.e., Mode 1 and Mode 2 (Fig. \ref{fig4}). Mode 1 is the non-reconfiguration mode, in which all the CS pairs of a reconfigurable VDPE are switched OFF, so that the reconfigurable VDPE operates like a regular VDPE to produce a VDP result of size \textit{N}. In contrast, Mode 2 is the reconfiguration mode, in which all the CS pairs of a reconfigurable VDPE are switched ON so that a total of \textit{y} in-parallel VDP results can be produced with each result being of size \textit{x}. We determine the re-aggregation size \textit{x} to be 9 because the DKV size \textit{s=9} is the most common, frequently used, smallest DKV size across various CNNs (Table \ref{dsc_tensor_info}). We reason that determining the value of \textit{x} based on the most common, smallest DKV size maximizes the opportunities for increasing processing parallelism and utilization efficiency. 

Since the reconfigurable VDPEs simply replace the regular VDPEs of the MAM and AMM TPCs, an RAMM/RMAM TPC is analogous in structure to an AMM/MAM TPC. From Fig. \ref{fig3ab} in Section III, an AMM/MAM TPC contains a VDPE block (containing multiple VDPEs), which is basically the structure that remains in the TPC if we mask off the DIV element of the TPC. Similarly, an RAMM/RMAM TPC would also typically contain a block of \textit{M} reconfigurable VDPEs. Since each reconfigurable VDPE is of size \textit{N}, such reconfigurable VDPE block would basically be of $M \times N$ dimensions. 

From Section II and III, the input kernel tensors are flattened into DKVs of size \textit{S} and then mapped onto the VDPE block of a TPC for processing. In that vein, to further explain our method of mapping for RAMM/RMAM TPCs, typically, a matrix $F_{(H,S)}$, which has \textit{H} rows with each row containing a DKV of size \textit{S}, is mapped onto one or multiple $M \times N$ sized reconfigurable VDPE blocks. This matrix $F_{(H,S)}$ can be written as a set of a total of \textit{H} $1 \times S$ sized DKVs, i.e., $F_{(H,S)}$ = $\{DKV_{(1,S)}|1|, \{DKV_{(1,S)}|2|,...,DKV_{(1,S)}|H|\}$. Each individual DKV in this set is basically a flattened kernel tensor. When matrix $F_{(H,S)}$ is mapped onto reconfigurable VDPE blocks, this set of DKVs is mapped onto the individual reconfigurable VDPEs of the blocks. After this mapping, the individual reconfigurable VDPEs operate in either Mode 1 or Mode 2, depending on how the size \textit{S} of the DKVs compares with the size \textit{N} of the reconfigurable VDPEs, given that the VDPEs have the re-aggregation size \textit{x=9}.

We advocate for selecting the most appropriate mapping and mode of operation (from Mode 1 or Mode 2) that can maximize the MRR utilization and processing throughput of the RAMM/RMAM TPCs. We identify three cases for the relation among \textit{N}, \textit{S}, and \textit{x=9} that drive the selection of the appropriate mapping and mode of operation. The mappings and modes of operation for these three cases are illustrated in Fig. \ref{rvdpeoperation}. For Fig. \ref{rvdpeoperation}, we selected the example RAMM TPC architecture from Table \ref{Nvalues} with \textit{N=20} for BR = 3 Gbps. These illustrations are further explained below.

\underline{\textbf{Case 1, \textit{S}$>$\textit{N}$>$\textit{x}}}: For this case, the reconfigurable VDPEs operate in Mode 1. In this case, before mapping the DKV matrix $F_{(H, S)}$ on the reconfigurable VDPEs, matrix $F_{(H, S)}$ is divided into multiple slices along the dimension \textit{S}. Since \textit{S} can be written as \textit{S}=\textit{b$\times$N+c}, the matrix $F_{(H, S)}$ is divided into a total of \textit{b+1} slices, with \textit{b} slices of size (\textit{H}, \textit{N}) each and one slice of size (\textit{H}, \textit{c}). Hence, $F_{(H, S)}$ can be written as $\{F_{(H,N)}^1 + .. + F_{(H,N)}^b + F_{(H,c)}^1\}$, where $F_{(H,N)}^1$,.., $F_{(H,N)}^b$, and $F_{(H,c)}^1$ are the slices of matrix $F_{(H,S)}$. Here, '\textit{+}' represents the concatenation operator. Since matrix  $F_{(H,S)}$ can be written as $\{DKV_{(1,S)}|1|,.., DKV_{(1,S)}|H|\}$ (as discussed earlier), slicing of $F_{(H,S)}$ in turn means slicing of all the DKV components of $F_{(H,S)}$. Consequently, every component DKV $DKV_{(1,S)}|H|$ of $F_{(H,S)}$ is also divided into \textit{b+1} slices along the dimension \textit{S}. Hence, $DKV_{(1,S)}|H|$ can be written as $\{DKV_{(1,N)}^1|H| + ..+ DKV_{(1,N)}^b|H| + DKV_{(1,c)}^1|H|\}$. Similarly, each DKV of matrix $F_{(H,S)}$ can be re-written as the concatenation of DKV slices. After the matrix $F_{(H,S)}$ has been sliced, each individual slice of every DKV of $F_{(H,S)}$ is mapped onto one reconfigurable VDPE running in Mode 1 operation. Each VDPE generates a partial VDP result corresponding to the mapped DKV slice. The partial VDP results from all the DKV slices that originated from a single original DKV (a component of $F_{(H,S)}$) are then accumulated at the \textit{psum} reduction network (not shown in the figure), to generate the final VDP result. Case 1 is shown in Fig. \ref{rvdpeoperation}(a), where input matrix $F_{(H,S)}$ with \textit{H=1} and \textit{S=32} is sliced based on \textit{b=1}, \textit{c=12}. The consequently generated individual DKV slices are then mapped on to reconfigurable VDPE1 and VDPE2 operating in Mode 1. The two VDPEs produce two partial VDP results $VDP_{(1, 20)}^{1}|1|$ and $VDP_{(1,12)}^{1}|1|$. These partial VDP results are then summed together.

\underline{\textbf{Case 2, \textit{N}$>$\textit{S}$>$\textit{x}}}: For this case, the reconfigurable VDPEs operate in Mode 2. In this case, before mapping the DKV matrix $F_{(H, S)}$ onto the reconfigurable VDPEs, matrix $F_{(H, S)}$ is divided into multiple slices along the dimension \textit{S}. Since \textit{S} can be written as \textit{S}=\textit{b$\times$x+c} (the values \textit{b} and \textit{c} here are often different from Case 1), the matrix $F_{(H, S)}$ is divided into a total of \textit{b+1} slices, with \textit{b} slices of size (\textit{H}, \textit{x}) each and one slice of size (\textit{H}, \textit{c}). Hence, $F_{(H, S)}$ can be written as $\{F_{(H,x)}^1 + .. + F_{(H,x)}^b + F_{(H,c)}^1\}$, where $F_{(H,x)}^1$,.., $F_{(H,x)}^b$, and $F_{(H,c)}^1$ are the slices of matrix $F_{(H,S)}$. Here, '\textit{+}' represents the concatenation operator. Since matrix $F_{(H,S)}$ can be written as $\{DKV_{(1,S)}|1|,.., DKV_{(1,S)}|H|\}$ (as discussed earlier), slicing of $F_{(H,S)}$ in turn means slicing of all the DKV components of $F_{(H,S)}$. Consequently, every component $DKV_{(1,S)}|H|$ of $F_{(H,S)}$ is also divided into \textit{b+1} slices along the dimension \textit{S}. Hence, $DKV_{(1,S)}|H|$ can be re-written as $\{DKV_{(1,x)}^1|H| + ..+ DKV_{(1,x)}^b|H| + DKV_{(1,c)}^1|H|\}$. After the matrix $F_{(H,S)}$ has been sliced, the total of \textit{H} DKV slices that belong to every matrix slice $F_{(H,x)}^1$ are mapped onto a total of \textit{(H/y)} different reconfigurable VDPEs, where each reconfigurable VDPE processes a total of \textit{y} slices (\textit{$y = N\geq2x  ?  floor(N/9) : 0$}). Hence, each reconfigurable VDPE in this case produces a total of \textit{y} VDP results in parallel, which basically renders a \textit{y$times$} throughput improvement for each reconfigurable VDPE, compared to Mode 1 of operation. In this case, once a given matrix slice has been mapped onto the total of \textit{(H/y)} VDPEs, the partial VDP results are produced for all DIVs from the input CNN layer that correspond to the mapped matrix slice in a stationary weight dataflow, before the next matrix slice is mapped for a new pass of all the DIVs. The partial VDP results from all the DKV slices that originated from a single original DKV (a component of $F_{(H,S)}$) are then accumulated at the \textit{psum} reduction network, to generate the final VDP result. Case 2 is shown in Fig. \ref{rvdpeoperation}(b), where input matrix $F_{(H,S)}$ with \textit{H=2} and \textit{S=16} is sliced based on \textit{b=1}, \textit{c=7}. The consequently generated individual DKV slices are then mapped on to a single VDPE operating in Mode 2 in two passes. Pass 1 generates $VDP_{(1, 9)}^{1}|1|$ and $VDP_{(1,9)}^{1}|2|$, Whereas Pass 2 generates $VDP_{(1, 7)}^{1}|1|$ and $VDP_{(1,7)}^{1}|2|$. The partial VDP results from Pass 1 are then summed together with respective partial VDP results from Pass 2 to generate two final VDP results.
    
\underline{\textbf{Case 3, \textit{N}$>$\textit{S}$\leq$\textit{x}}}:
For this case, the reconfigurable VDPEs operate in Mode 2. The input matrix $F_{(H, S)}$ is directly mapped onto the individual reconfiguration VDPEs, after writing it in terms of DKVs as $F_{(H,S)} = \{DKV_{(1,S)}|1|,.., DKV_{(1,S)}|H|\}$. Here, H DKVs corresponding to matrix $F_{(H,S)}$ are mapped onto a total of (\textit{H}/\textit{y}) reconfigurable VDPEs operating in Mode 2, where each VDPE processes a total of \textit{y} DKVs in parallel. Case 3 is shown in Fig. \ref{rvdpeoperation}(c), where input matrix $F_{(H,S)}$ with \textit{H=2} and \textit{S=8} is mapped on to a reconfigurable VDPE operating in Mode 2. The VDPE processes \textit{y=2} DKVs, generating \textit{y=2} VDP final results $VDP_{(1,8)}|1|$ and $VDP_{(1,8)}|2|$.

\underline{\textbf{Discussion:}} Compared to Mode 1, Mode 2 operation of the reconfigurable VDPE increases the hardware/MRR utilization, while simultaneously increasing the processing throughput by up to \textit{y$\times$}. This throughput improvement makes it tolerable to have the extra area overhead of additional \textit{y} CS pairs per reconfigurable VDPE for Mode 2 operation. It can be argued that \textit{y$\times$} improvement in throughput can also be achieved by employing \textit{y$\times$} more VDPEs (without CS pairs) in Mode 1 operation. However, employing \textit{y$\times$} more VDPEs in Mode 1 operation incurs the area overhead of additional \textit{N$\times$y} MRRs (as each VDPE has \textit{N} MRRs), whereas employing \textit{y} CS pairs in one reconfigurable VDPE incurs the area overhead equivalent to additional \textit{6$\times$y} MRRs (as the area of 1 CS pair = area of 6 MRRs). Since for all TPCs from Table \ref{Nvalues}, \textit{6$\times$y} is less than \textit{N$\times$y}, Mode 2 operation turns out to be highly efficient than Mode 1 operation. As for modern CNNs, more than 40\% of the DKVs (Table \ref{dsc_tensor_info}) belong to Case 2 (\textit{N}$>$\textit{S}$>$\textit{x}) and Case 3 (\textit{N}$>$\textit{S}$\leq$\textit{x}) above, the substantially improved efficiency for Mode 2 operation also translates in up to 78.2\% and 54.71\% higher VDPE utilization, respectively, for our RAMM and RMAM TPCs in Fig. \ref{vdpe_utilization}, compared to their respective baseline AMM (DEAPCNN) and MAM (HOLYLIGHT) TPCs.



\subsection{Design of MRR Comb Switches}
Compared to the modulation MRRs of the DIV and DKV elements (Fig. \ref{fig3ab}), our utilized MRR-based CSs (Fig. \ref{fig4}) typically have a larger ring radius \cite{combswitch}. From Section V.A, a CS has the capability of filtering a comb of \textit{x} wavelength channels from the incoming \textit{N} wavelength channels. To achieve this capability, the periodic resonance passbands of the CS need to overlap with this comb of \textit{x} distinct wavelength channels. Generally, the resonance passband of a modulation MRR periodically repeats at the spectral distance known as Free Spectral Range (FSR), and two adjacent wavelengths in the incoming comb of \textit{N} wavelengths have a channel spacing of $\Delta$ (Eq. \ref{eq3}).
\begin{equation}
    \Delta = \frac{FSR}{N+1}
\label{eq3}
\end{equation}
Therefore, to filter \textit{x} distinct wavelengths from these \textit{N} incoming wavelengths separated by $\Delta$, the FSR of the CS needs to be equal to $CS_{FSR}$ (Eq. \ref{eq4}). 

\begin{equation}
    CS_{FSR} = \frac{\textit{N}*\Delta}{x}
\label{eq4}
\end{equation}

Since \textit{N} varies for our RAMM and RMAM TPCs across different BRs (Table \ref{Nvalues}), the required FSR for the CSs would also vary across different  BRs. The FSR for a CS can be defined by appropriately defining its radius. Therefore, we designed the CSs with desired FSRs required for our RAMM and RMAM TPCs for various BRs by appropriately defining the radius of a primitive MRR. For that, we used  the photonics foundry-validated MODE and INTERCONNECT tools from  Ansys/Lumerical \cite{18}. Table \ref{COMB_SWITCH_INFO} lists the design parameters of our designed CSs. From Table \ref{COMB_SWITCH_INFO}, a CS can incur insertion losses. This can impact the achievable \textit{N} for a TPC. Our scalability analysis in Section \ref{sec3b} accounts for this fact.

\subsection{System Level Implementation} \label{accelerator overview}
\begin{figure}[h]
  \centering
  \includegraphics[scale = 0.4]{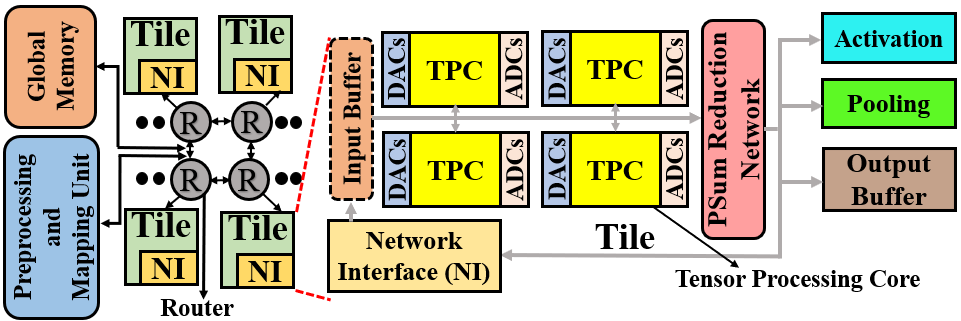}
  \caption{System level overview of a CNN accelerator that employs our RMAM/RAMM TPCs.}
  \label{fig1}
\end{figure}

Fig. \ref{fig1} illustrates the system level implementation of our accelerators. It consists of a global memory for storing CNN parameters, a pre-processing and mapping unit for decomposing the tensors into DIVs/DKVs and mapping them onto the DIV/DKV elements. It has a mesh of tiles connected to routers and this mesh network facilitates parameter communication among tiles. Each tile consists of 4 RMAM/RAMM TPCs interconnected (via H-tree network) with output buffer, activation and pooling units. Due to their analog nature, the RMAM/RAMM TPCs require DACs and ADCs as well. In addition, each tile also contains \textit{psum} reduction network to enable summing up of the intermediate \textit{psum}s. Depending on the type of convolution operations being processed (SC/DC/PC) and their tensor sizes (Table \ref{dsc_tensor_info}), the mapping unit sends control signals to the individual RAMM/RMAM TPCs to reconfigure their operational modes (Mode 1/Mode 2).

\medskip

\begin{table}[H]
\centering

\caption{Design parameters of various comb switch (CS)
designs used in our RMAM and RAMM TPCs for various BRs.}
\label{COMB_SWITCH_INFO}
\begin{tabular}{|cccc|}
\hline
\multicolumn{1}{|c|}{\textbf{Bit Rate (BR) (Gbps)}}           & \multicolumn{1}{c|}{\textbf{1}}  & \multicolumn{1}{c|}{\textbf{3}} & \textbf{5}  \\ \hline \hline
\multicolumn{4}{|c|}{\textbf{RAMM TPC}}                                                                                                \\ \hline
\multicolumn{1}{|c|}{\textbf{\textit{N}}}                   & \multicolumn{1}{c|}{31}          & \multicolumn{1}{c|}{20}          & 16          \\ \hline
\multicolumn{1}{|c|}{\textbf{$CS_{FSR}$}}               & \multicolumn{1}{c|}{4.83nm}      & \multicolumn{1}{c|}{5 nm}        & NA          \\ \hline
\multicolumn{1}{|c|}{\textbf{Radius}}              & \multicolumn{1}{c|}{18.17 $\mu$m}     & \multicolumn{1}{c|}{17.5 $\mu$m}      & NA          \\ \hline
\multicolumn{1}{|c|}{\textbf{No of CS Pairs}}      & \multicolumn{1}{c|}{3}           & \multicolumn{1}{c|}{2}           & 0           \\ \hline
\multicolumn{1}{|c|}{\textbf{Insertion Loss (dB)}} & \multicolumn{1}{c|}{0.029}       & \multicolumn{1}{c|}{0.028}       & 0           \\ \hline
\multicolumn{4}{|c|}{\textbf{RMAM TPC}}                                                                                                \\ \hline
\multicolumn{1}{|c|}{\textbf{\textit{N}}}                   & \multicolumn{1}{c|}{\textbf{43}} & \multicolumn{1}{c|}{\textbf{28}} & \textbf{22} \\ \hline
\multicolumn{1}{|c|}{\textbf{$CS_{FSR}$}}               & \multicolumn{1}{c|}{4.65 nm}     & \multicolumn{1}{c|}{5.35nm}      & 4.54 nm     \\ \hline
\multicolumn{1}{|c|}{\textbf{Radius}}              & \multicolumn{1}{c|}{18.98 $\mu$m}    & \multicolumn{1}{c|}{16.2 $\mu$m}     & 19.49 $\mu$m    \\ \hline
\multicolumn{1}{|c|}{\textbf{No of CS Pairs}}      & \multicolumn{1}{c|}{4}           & \multicolumn{1}{c|}{3}           & 2           \\ \hline
\multicolumn{1}{|c|}{\textbf{Insertion Loss (dB)}} & \multicolumn{1}{c|}{0.029}       & \multicolumn{1}{c|}{0.026}       & 0.031       \\ \hline
\end{tabular}
\end{table}

\section{Evaluation}\label{Evaluation}
\subsection{Simulation Setup}
To evaluate our designed RAMM and RMAM accelerator architectures, we simulated the inference of various depthwise separable convolutions based CNNs such as EfficientNetB7 \cite{efficientnet}, Xception \cite{xception}, NASNetMobile\cite{NASNetmobile}, and ShuffleNetV2 \cite{shufflenet} with input batch size of 1. These CNNs comprehensively cover wide variations seen in CNNs in terms of tensor sizes. We developed a custom, transaction-level, cycle-true python-based simulator to model CNN inference on MRR-based TPC accelerators with weight-stationary dataflow. 

\smallskip

\begin{table}[H]
\centering

\caption{ADC area and power overheads.}
\label{adc}
\begin{tabular}{|c|c|c|}
\hline
\textbf{ADC}     & \textbf{Area (mm$^2$)} & \textbf{Power} \\ \hline
1 Gbps \cite{adc1gbps}  & 0.002               & 2.55 mw        \\ \hline
3 Gbps \cite{adc3gbps}  & 0.021               & 11mw           \\ \hline
5 Gbps \cite{adc5gbps} & 0.103               & 29 mw          \\ \hline
\end{tabular}
\end{table}

We compared our RAMM and RMAM accelerator architectures with the baseline AMM (DEAPCNN \cite{15}), MAM (HOLYLIGHT \cite{17})  and the latest variant of AMM design (CROSSLIGHT \cite{14}). We evaluate these accelerators at 4-bit precision and across different BRs such as 1 Gbps, 3 Gbps, and 5 Gbps. The \textit{N} values corresponding to different BRs are taken from Table \ref{Nvalues} for our system level analysis. Table \ref{adc} and Table \ref{table3} give the parameters used for evaluating the overheads of the peripherals. We consider each laser diode to emit input optical power of 10 mW (10 dBm) (Table \ref{abbrevations}) \cite{15}. Multiplexer and splitter parameters are taken from \cite{17}, and other VDP element parameters are listed in Table \ref{table4}.

We performed area proportionate (AP) analysis, for which, we altered the VDPE count of each accelerator so that the accelerators' area matched with the area of the RMAM accelerator with VDPE count of 512. Table \ref{VDPECount} reports our obtained area-proportionate VDPE counts for all our considered accelerators. We evaluate the metrics such as Frames Per Second (FPS) and FPS/W (energy efficiency) for our considered accelerator architectures.

\medskip

\begin{table}[H]
\centering
\caption{Accelerator Peripherals Parameters \cite{17}.}
\label{table3}
\begin{tabular}{|c|c|c|c|}
\hline
\textbf{}                  & \textbf{\begin{tabular}[c]{@{}c@{}}Power(mW)\end{tabular}} & \textbf{\begin{tabular}[c]{@{}c@{}}Area($mm^2$)\end{tabular}} & \textbf{Latency}   \\ \hline
\textbf{DAC \cite{dac10gbps}}               & 30                                                         & 0.034                                                           & 0.78ns             \\ \hline
\textbf{Reduction Network} & 0.05                                                           & 0.03E-3                                                        & 3.125ns            \\ \hline
\textbf{Activation Unit}   & 0.52                                                           & 0.6E-3                                                         & 0.78ns             \\ \hline
\textbf{IO Interface}      & 140.18                                                         & 24.4E-3                                                        & 0.78ns             \\ \hline
\textbf{Pooling Unit}      & 0.4                                                            & 0.24E-3                                                        & 3.125ns            \\ \hline
\textbf{eDRAM }    & 41.1                                                           & 166E-3                                                         & 1.56ns             \\ \hline
\textbf{Bus}               & 7                                                              & 9E-3                                                           & 5 cycles\\ \hline
\textbf{Router}               & 42                                                              & 0.151                                                          & 2 cycles \\ \hline
\end{tabular}
\end{table}

\smallskip

\begin{table}[H]
\centering
\caption{VDP Element Parameters \cite{14}.}
\label{table4}
\begin{tabular}{|c|cc|}
\hline
\textbf{DKV/DIV MRR Q-factor} & \multicolumn{2}{c|}{8000}                                   \\ \hline
\textbf{DKV/DIV MRR FWHM}     & \multicolumn{2}{c|}{0.2 nm}                                 \\ \hline
\textbf{Sensitivity of PD}    & \multicolumn{2}{c|}{-20 dBm}                                \\ \hline
                              & \multicolumn{1}{c|}{\textbf{Power (mW)}} & \textbf{Latency} \\ \hline
\textbf{EO Tuning}            & \multicolumn{1}{c|}{80 $\mu$W/FSR}       & 20 ns            \\ \hline
\textbf{TO Tuning}            & \multicolumn{1}{c|}{27.5 mW/FSR}         & 4 $\mu$s          \\ \hline
\textbf{TIA}                  & \multicolumn{1}{c|}{7.2 mW}              & 0.15 $\mu$s      \\ \hline
\textbf{Photodetector}        & \multicolumn{1}{c|}{2.8 mW}              & 5.8 ps           \\ \hline
\end{tabular}
\end{table}

\smallskip
\begin{table}[H]
\centering

\caption{VDPE counts of various accelerators.}
\label{VDPECount}
\begin{tabular}{|c|lll|}
\hline
\multirow{2}{*}{\textbf{Accelerators}}                   & \multicolumn{3}{c|}{\textbf{BR   (Gbps)}}                                                           \\ \cline{2-4} 
                                    & \multicolumn{1}{c|}{\textbf{1}} & \multicolumn{1}{c|}{\textbf{3}} & \multicolumn{1}{c|}{\textbf{5}} \\ \hline
\textbf{RMAM}                       & \multicolumn{1}{l|}{512}        & \multicolumn{1}{l|}{512}        & 512                             \\ \hline
\textbf{RAMM}                       & \multicolumn{1}{l|}{587}        & \multicolumn{1}{l|}{576}        & 567                             \\ \hline
\textbf{MAM (HOLYLIGHT   \cite{17})} & \multicolumn{1}{l|}{568}        & \multicolumn{1}{l|}{562}        & 547                             \\ \hline
\textbf{AMM (DEAPCNN   \cite{15})}   & \multicolumn{1}{l|}{656}        & \multicolumn{1}{l|}{629}        & 620                             \\ \hline
\end{tabular}
\end{table}


\subsection{Evaluation Results}

\begin{figure}[H]
  \centering
  \includegraphics[width=\linewidth]{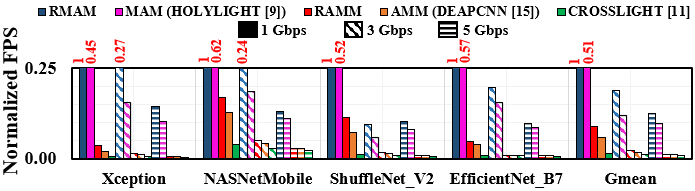}
  \caption{Area proportionate comparison of FPS for various accelerators across different CNNs and bit rates (BRs). Results are normalized with respect to RMAM at 1 Gbps.}
  \label{fig10}
\end{figure}

Fig.\ref{fig10} shows the FPS results for various accelerators at different BRs, normalized to RMAM at 1 Gbps. Our RMAM accelerator on gmean outperforms MAM (HOLYLIGHT), AMM (DEAPCNN), and CROSSLIGHT for all BRs, such as 1 Gbps, 3 Gbps, and 5 Gbps. At 1 Gbps, RMAM achieves 1.8$\times$, 17.1$\times$, and 65$\times$ better FPS than MAM (HOLYLIGHT), AMM (DEAPCNN), and CROSSLIGHT, respectively, on gmean across the CNNs. From Section \ref{sec3b}, the \textit{N} values decrease with increase in BR, which leads to lower throughput with increase in BR for various accelerators. For instance, RMAM's FPS drops by 5.3$\times$ and 8$\times$ at 3 Gbps and 5 Gbps respectively, compared to its FPS at 1 Gbps. Therefore, compared to the 3-Gbps (5-Gbps) variants of MAM (HOLYLIGHT), AMM (DEAPCNN), and CROSSLIGHT, respectively, our 1-Gbps RMAM variant achieves 8.3$\times$ (10.2$\times$), 52.57$\times$ (79.8$\times$), and 86$\times$ (106$\times$) better FPS. These FPS benefits are because of our RMAM variants' higher throughput and more efficient processing through reconfiguration (Mode 2 operation). Our other accelerator RAMM, at 1 Gbps, achieves 1.54$\times$ and 5.8$\times$ better FPS compared to AMM (DEAPCNN) and CROSSLIGHT, respectively. Even at higher BRs, RAMM performs better than AMM (DEAPCNN) and CROSSLIGHT. However, at 5 Gbps, because of low \textit{N}, reconfiguration is not supported in RAMM due to the condition \textit{$y = N\geq2x  ?  floor(N/9) : 0$}. Therefore RAMM is exactly identical to AMM (DEAPCNN) at 5 Gbps. Overall, our RMAM accelerator gives better FPS compared to other accelerators across all BRs.

\begin{figure}[H]
  \centering
  \includegraphics[width=\linewidth]{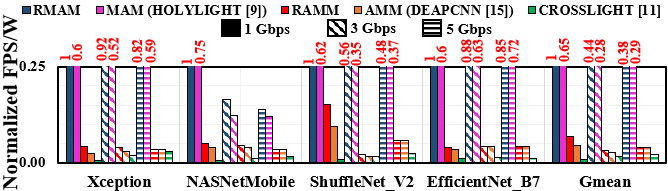}
  \caption{Area proportionate comparison of FPS/W for various accelerators across different CNNs and bit rates (BRs). Results are normalized with respect to RMAM at 1 Gbps.}
  \label{fig11}
\end{figure}

Fig. \ref{fig11} shows the FPS/W (energy efficiency) results for various accelerators across different BRs, normalized to RMAM at 1 Gbps. Our RMAM and RAMM accelerators also achieve better FPS/W compared to the other accelerators. At 1 Gbps, RMAM achieves 1.5$\times$, 27.2$\times$, and 171$\times$ better FPS/W than MAM (HOLYLIGHT), AMM (DEAPCNN), and CROSSLIGHT, respectively, on gmean across various CNNs. Similar to the FPS results, when compared to the 3-Gbps (5-Gbps) variants of MAM (HOLYLIGHT), AMM (DEAPCNN), and CROSSLIGHT, our 1-Gbps RMAM variant achieves 4.2$\times$ (4$\times$), 46.4$\times$ (29.6$\times$), and 80.7$\times$ (54$\times$) better FPS/W, respectively. These energy efficiency benefits are also because of the improved VDPE utilization achieved from the superior reconfigurability of our RMAM accelerator. The improved VDPE utilization better amortizes the static power consumption of the constituent hardware components (e.g., MRRs, CSs, ADCs, DACs) to yield better energy efficiency. Our other accelerator RAMM, at 1 Gbps, achieves 1.5$\times$ and 9.7$\times$ better FPS/W compared to AMM (DEAPCNN) and CROSSLIGHT, respectively. Even at higher BRs, RAMM performs better than AMM (DEAPCNN) and CROSSLIGHT. However, at 5 Gbps, RAMM and AMM (DEAPCNN) have equal energy efficiency. Overall, our RMAM provides better energy efficiency compared to the other accelerators across different BRs.

\section{Conclusion}

In this paper, we presented the use of our invented reconfigurable VDPEs as a novel way of introducing flexibility in the photonic MRR-based CNN accelerators. Our reconfigurable VDPEs employ a set of comb switches to enable dynamic maximization of the size compatibility between the VDPEs and the CNN tensors that are processed using the VDPEs. We then used our reconfigurable VDPEs to enhance the MRR-based CNN accelerators of the AMM and MAM categories. Consequently, we derived different variants of Reconfigurable MAM (RMAM) and Reconfigurable AMM (RAMM) accelerators that operate at different bit rates. We evaluated different variants of our RMAM and RAMM accelerators against three prior works, for the inference of four modern CNNs with mixed-sized tensors. Our evaluation indicates that our RMAM and RAMM accelerators are significantly better at striking a balance between the hardware utilization and CNN processing latency, which in turn provides them with substantial improvements in FPS and FPS/W, with equal area consumption, compared to the photonic MRR-based accelerators from prior works. These results promote the use of our RMAM and RAMM accelerators for efficient processing of future CNNs having a wide variety in their employed tensor sizes. 

\section*{Acknowledgments}
We thank the anonymous reviewers whose valuable feedback helped us improve this paper. We would also like to acknowledge the National Science Foundation (NSF) as this research was supported by NSF under grant CNS-2139167.


\bibliographystyle{IEEEtran}
\bibliography{references}

\vfill

\end{document}